# Biologically Inspired Execution Framework for Vulnerable Workflow Systems

Sohail Safdar, Mohd. Fadzil B. Hassan, Muhammad Aasim Qureshi, Rehan Akbar

Department of Computer & Information Sciences,
Universiti Teknologi PETRONAS,
Malaysia

.

*Abstract*—The main objective of the research is to introduce a biologically inspired execution framework for workflow systems under threat due to some intrusion attack. Usually vulnerable systems need to be stop and put into wait state, hence to insure the data security and privacy while being recovered. This research ensures the availability of services and data to the end user by keeping the data security, privacy and integrity intact. To achieve the specified goals, the behavior of chameleons and concept of hibernation has been considered in combination. Hence the workflow systems become more robust using biologically inspired methods and remain available to the business consumers safely even in a vulnerable state.

## I. INTRODUCTION

Now days, the world is moving towards the economic growth, achieving the business goals is of prime importance. The major requirement for achieving the business goals is the reliable business processes to provide customers with great deal of satisfaction in terms of Quality of Services. Customized software are in common use to provide solutions for different business processes to increase the performance and providing quick, in time concrete trade results. These business processes are known as business workflow processes or business workflows in terms of computing.

Workflow Management Systems (WFMS) are the systems that are used to automate, manage, monitor and control the execution of the workflow processes. Workflow process is a business process for an enterprise. Workflow process contains set of workflow activities that are required to be completed in the specified sequence to finish the workflow process. Each workflow activity is a single or set of instructions to be executed.

Workflows are currently very active area of research. Various efforts have been made to provide the business process optimizations and improving the quality of services. Improving coordination among cooperating workflows, process synchronizations, robustness of operational workflows, workflow representational techniques and secure workflows are all very hot areas in which lots of research work is going on.

The major concern of any business is to secure all its data hence to keep customers' as well as company's privacy intact. The customer's satisfaction in terms of getting good quality services well in time along with the guarantee of protected and secured transactions are of prime importance.

Hence various mechanisms have been provided over the period of time to provide secured workflow transactions using Workflow transaction management and IDS (Intrusion Detection System).

The current research is motivated from the efforts that have been made to provide secure workflow systems and the problems associated with those systems. Currently, the WFMS rely merely on IDS for intrusion detection. Once intrusion is detected the whole system is set to wait state and the running process is undo and redoes to recover the faulty parts. This practice might lead to lose the customer satisfaction to use the system as customers always willing to have the timely and accurate result with all of the protection provided. So when the system is in vulnerable state then the questions arise

How data would be secured and its privacy would be maintained, when intrusion is found?

How can the current and the remaining activities safely continue their execution?

How will the workflow engine be able to execute workflow process in a robust fashion and ensures the secure availability of the system along with the integrity of data to all the customers?

All the above mentioned questions concludes the problem associated to the workflow systems that are in vulnerable state due to some intrusion detected during their execution. The problem is to avoid the possibility of system to enter into wait state whenever the intrusion is detected. The current research is dealing with all of the concerns associated with the problem to provide best possible solution. Specifically, the problem statement for the research is:

In the case of intrusion threat, the system goes into unsafe state. The workflow management system should ensure in time availability of services while keeping the data integrity intact and continue the workflow process robustly to provide satisfactory results to the end user/customer.

The main objective of this research is to design a framework that will provide the data and services availability all the time by keeping data security and privacy intact, when the intrusion strikes and system goes into unsafe state. The proposed framework will be utilizing the biological inspired mechanisms to provide data protection, security and privacy.

The following section will explain the background of the related literature in the context proposed research area followed by the overview of the proposed research. The details of the related concepts and proposed frame work will be explained in the proposed methodology.

## II. BACKGROUND

WFMS is a very hot area of research. Various efforts have been made in areas of workflow representations, adaptive workflows, workflows performance and management issues, workflows security and self healing in workflows. The current research is also related to the area of security and workflow system recovery. Various existing work involves different approaches for intrusion detection and then system recovery. Multi-Version Objects [1] approach is to replace the dirty objects with the clean version to recover the system, and the whole system works in more than one version of each object. Whenever there is an intrusion that infects the data object, the system is stopped and then recovered to the previous state with the help of these clean versions of the objects. The graph theory in theoretical computer science is also referred while recovering procedures are applied [18], [19], [20]. Trace back recovery [2] mechanism is based on Flow-Back Recovery Model [16] that uses the traces of the flow of execution and then recovers the workflow system. Another approach utilizes the workflow specification to detect the intrusions with the help of independent Intrusion Detection System. It proposes an "Attack Tree Model" [3] to describe the major goal of the attack and then splitting it to the sub goals. The work focuses to provide the system recovery through dynamic regeneration of workflow specification. The Undo and Redo mechanism is utilized to recover and bring the system to consistent state. This approach deals with the exception raised by the intrusions and regenerate the workflow specification dynamically for the workflow to execute successfully. Architecture consists of BPEL (Business Process Enterprise Language) Engine and Prolog Engine for intelligence is utilized to regenerate the workflow dynamically [3]. There is another architecture named MANET [4] that provides additional features of Mobile services, Workflow Modeler and policy Decision point to regenerate the Workflow specification more effectively [4]. Vulnerabilities are also detected by the use of a workflow layer on any system as a non intrusive approach is proposed based on this architecture for survivability in the cyber environment [5]. The overall security is based on the model [6] that Threat Agent causes threats that cause vulnerability. Vulnerability that causes risks can be reduced by a safe guard that protects an asset [6].There are different approaches like Do-It-All-Up-Front Approach, All or Nothing, Threat Modeling and Big Bang approach etc. for ensuring security on web and has their own pros and cons [7]. Ammann et al. [11] deals with the transactions done by malicious users and recover the system by cleaning the infected data items due to these transactions and hence undo all those transactions. Panda et al. [17] provides number of algorithms to recover the system based on the dependency information that is stored separately. Eder and Liebhart [14] also studies potential failures in workflows and found its possible recovery mechanisms. Problems associated with recovery and rollback in distributed environment has also been handled [15]. Few more work related concurrency control in databases and its transaction [12] [13]. It must be noted that whenever an intrusion strikes a workflow system and is detected, the system must be stopped immediately to avoid any data infection for maintaining its integrity. So all of the recovery method needs the mechanism of undo all the faulty areas that require the system to wait and then start the process again to redo things once the system gets back to safe state. But making the system wait for the recovery and redoing the same processing again annoys the customer from the system. Hence ensuring the availability of the system even in the unsafe state is very much required that has not addressed yet by anyone.

## III. OVERVIEW OF THE PROPOSED RESEARCH

### A. Problem Statement

Business requires 100% availability of their workflow systems, so that the services have been provided to the customers securely and the customers have 100% satisfaction on their services. In the case of intrusion detected, the system needs to be stop so that it can be recovered from the possible threat. Due to which the availability of services at that time might not be possible. Hence whenever a system goes in to unsafe state due to some intrusion, the workflow management system should provide.

- Security and privacy of data

- In time availability of correct data to ensure the completion of desired transaction.

- Complete the workflow process robustly to provide satisfactory services to the end user/customer.

### B. Objectives

The main objectives of the research are following.
Design an alternative execution framework for the workflows in vulnerable state such that it

- Provides the robust execution of the entire workflow process.

- Ensure the data security and privacy.

- Availability of in time correct data to the customers.

### C. Concerns

There are certain concerns associated with the methodology to achieve the objectives. How data can be secured and its privacy can be maintained, when intrusion is found? How can the remaining activities continue their execution? The argument for the first concern lies in the concept of chameleon characteristics, hence we can say that by applying chameleon characteristics to database portion for the specific ongoing activity to carry on. However the argument for the second concern is that we can apply the concept of data hibernation.

Following section will provide the definitions of the concepts of Chameleon data sources and Data hibernation.

### D. Definitions

The following are the definitions of the useful concepts regarding the research paper.

#### 1) Chameleon Data Sources

The term is taken from the concept of chameleon characteristics of changing color and is defined as changing of data values to unreadable data when the data source is found to be under threat. The concept is shown in Figure 1 and Figure 2.

#### 2) Data Hibernation

The term is driven from the concept of hibernation in animals in which animals go to sleep for a certain period of time under soil and is defined as shifting a data from the original data source to multiple dimensions when there is a threat to its integrity and return back to original source when the threat is removed. The concept is shown in Figure 3 and Figure 4.

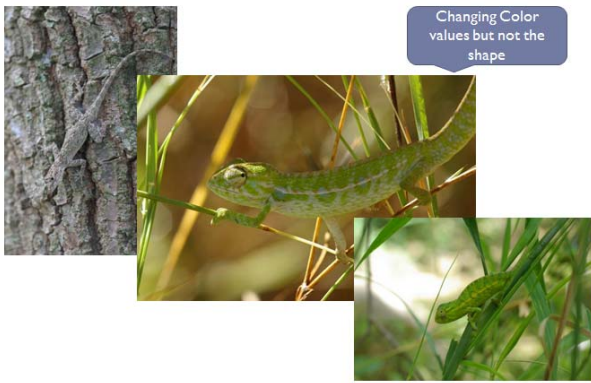

Figure 1. Chameleon and its Characteristics

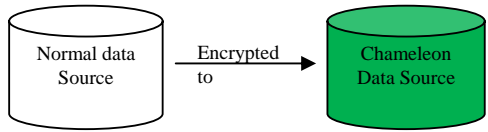

Figure 2. Chemeleon Data Source Behavior

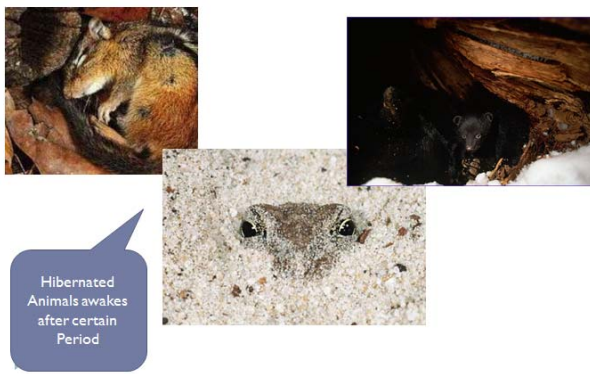

Figure 3. Behavior of Animal Hibernation

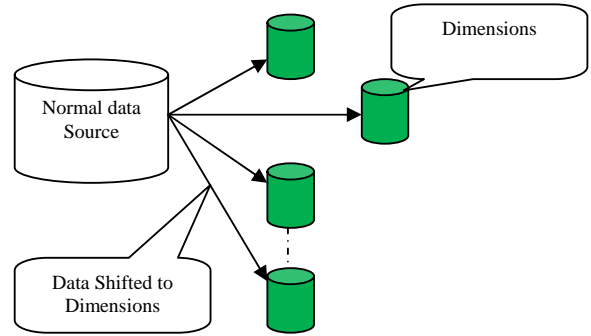

Figure 4. Behavior of Data Hibernation

### IV. PROPOSED METHODOLOGY

Proposed methodology is the base line for the desired framework to provide the execution of vulnerable workflows to provide services and data availability. The methodology includes designing a mechanism that provides and ensures the data security, integrity and privacy in the operational workflows. There is also a requirement of a mechanism to make the data available to the customer retaining its integrity when the system is in unsafe state. These two mechanism leads to the proposed framework for the execution of the vulnerable workflow system. The following is the explanation of the proposed methodology.

### A. Explanation

#### 1) Designing a mechanism, to provide and ensure the data security and privacy in operational workflows:

This whole mechanism is biologically inspired from the behavior of changing colors such as Chameleon does and hibernation mechanism in the wild life. There are two milestones to achieve while dealing with this issue. One is handling ongoing activity while other is handling the upcoming activities.

Handling the ongoing activity while the system is declared as unsafe due to some intrusion requires an implementation of Chameleon Data Sources concept as follows:

##### a) Role of Chameleon Data Sources

The concept is drawn from the natural phenomenon of changing colors by Anole and Chameleon in case of any threat to provide them an appropriate sabotage. Getting inspiration from this concept and applying it to the portion of data source that has been utilized by the ongoing activity leads the data to be sabotage and becomes secured from the threat of intrusion and hence keep its privacy and integrity. Applying the concept requires data in the database should be changed dynamically from the meaningful state into meaningless state by using the encryption rules. It is not only the encryption of data but it is dynamically applying the encryption to the data sets whenever data's privacy seems to be in danger.

Simultaneously, the data associated with the upcoming activities should also be handled, which can be done by applying the data hibernation concept as following

### b) Role of Data Hibernation

The concept is also raised from the animal behavior to reside under soil for the specific time. During Modeling Phase, the chunks of larger database needs to be divided and modeled correctly, so that it can easily be integrated with the larger database in terms of moving the portion of unsafe data to that small dimension and shifting them back when the system returns in a safe state again.

The portion of database schema, whose data needs to be hibernating, should be changed using dimensional modeling. Each dimension of database is one that is referred during any specific workflow activity execution, i.e. the dimension is w.r.t the context of workflow activity. Each dimension is a normalized dimension unlike the dimensions in the data warehousing context. The data is then transformed using ETL into that dimension and needs to be accessed from that area until the system regain its safe state.

### 2) Designing a mechanism to make the data available in its correct form to the customer even if the system is in unsafe state:

Dealing with the ongoing activity requires continue referring the same portion of the database on which the current transaction is based on. Applying the dynamic encryption to that portion of database making it a chameleon natured will help to solve the problem in the current scenario. Not only data becomes meaningless for all the external sources but also it becomes ready to use by the alternative commands that can be able to decrypt it and use it. The point of consideration here is to make the portion of that database as read only so that encrypted data might not be overwritten by intrusion activity with dirty data to become useless at all. Hence by doing so, the change that has been made by the ongoing activity should be stored using caching. Once the data is completed its required transformation then it should be written in the relevant hibernated dimension. All of the upcoming activities will refer the hibernated data from the respective dimensions.

### V. PROPOSED FRAMEWORK FOR ROBUST EXECUTION OF VULNERABLE WORKFLOWS

The following is the proposed algorithm for robust execution of the vulnerable workflows to provide data and services availability to the customers in a non discrete fashion.

1. The intrusion attack is detected by a workflow process using some IDS.

2. Workflow server signals the flag to the workflow engine.

3. On receiving the flag, the workflow engine interrupts the resource manager.

4. Resource manager forces the active data source to change its state and hibernate the data in all of the dimensions except that of the currently active data.

5. The current workflow activity accesses the data using encryption and decryption mechanism. However the upcoming workflow activities in the running system will access the data from the hibernated data source.

6. Using these two key phenomenon, workflow transaction will not stop and even in the unsafe state the whole system will robustly keep on operating in a secured, available and manageable fashion.

### A. Explanation

The above mentioned working has to be done while workflow system follows an alternative path due to the intrusion threat and needs to be carried until the system is recovered fully from the threat as shown in the Figure 5 and Figure 6.

In Figure 5, when the data in the main database has encrypted, at the same time, the data becomes read only so that all the activities to spoil the data by writing garbage on it can also be controlled. After the current active tasks finishes its execution then the transformed results in the memory and the existing data inside encrypted database portion would be written in the respective dimension. The other dimensions can be populated during the execution of the current task as background process.

When the system is recovered from the possible threat or the threat is rectified then the data in the dimensions will be transferred back to the original database at their appropriate location. This whole phenomenon can be seen in Figure 6.

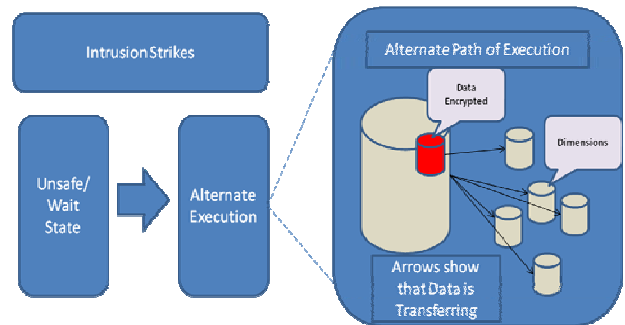

Figure 5. Workflow System state when intrusion strikes

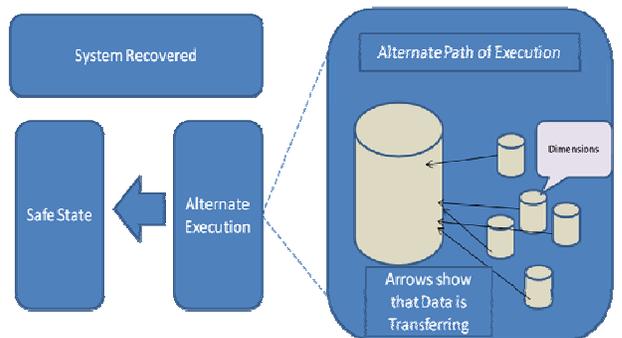

Figure 6. Workflow System state when intrusion is rectified

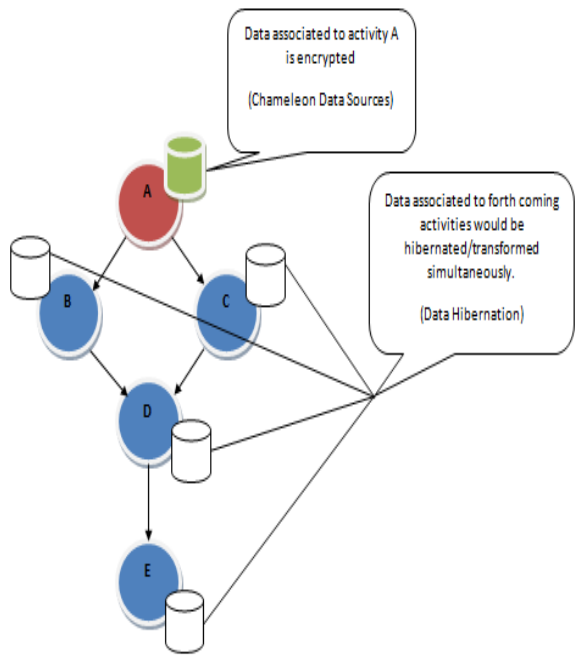

Figure 7: Overall view of the execution of vulnerable workflow in a secured fashion using the proposed framework

Figure 7 shows the overall view of the workflow process robustly executing under the proposed framework guidelines, providing the availability of the services and data to the customers.

### B. Strength & Weaknesses of the Proposed Framework

The framework provides the workflow with great strength to continue its execution robustly in a secured manner by making the availability of the data and services possible for the customers. Due to this robustness and security, the end users and customers rely on the system with more confidence. On the other hand the proposed framework is targeting the centralized data sources. Framework does not target the issues related to distributed data sources that has to be taken care as its future implications.

### CONCLUSION

The research contributes to resolve the issue of service unavailability to the end user or business customers in case of intrusion intervention in the workflow system. The services not only are available but in a secured fashion by keeping the privacy and integrity of the data intact. Moreover the research is a pioneer step in the area of making system keep working even in the unsafe state, so to provide maximum satisfaction to the customer. Providing such framework enables the enterprises to run their own customized solutions based on the provided guideline. The work also focuses to provide the workflow process providing self security. The framework is targeting the centralized data source, it may however be extended to cater the distributed data sources and services in future.